\documentclass[12pt]{article}
\usepackage{amsmath}
\usepackage{amsfonts}
\usepackage{euscript}
\usepackage{epsf}
\newfont{\smithscript}{rsfs10 at 12pt}
\newfont{\bsmscript}{rsfs5 at 12pt}
\newcommand{\mathst}[1]{\mbox{\smithscript #1}}

\newcommand{\mathbold}[1]{\mbox{\boldmath $#1$}}
\begin{document}
\vspace *{2cm} 
\begin{flushright}
UTEXAS-HEP-00-1
\end{flushright}

\begin{center}
{\Large \bf Ramond-Ramond Field Transformation}

\bigskip

Yungui Gong \footnote{Email: ygong@physics.utexas.edu} 

Physics Department, University of Texas at Austin, Austin, TX 78712, U.S.A.
 
\bigskip
 
\vskip 0.8in
{\Large\bf Abstract}
\end{center}
We find that the mixture of Ramond-Ramond
fields and Neveu-Schwarz two form are transformed
as Majorana spinors under the T-duality group $O(d,d)$.
The Ramond-Ramond field transformation under
the group $O(d,d)$ is realized in a simple
form by using the spinor representation. The 
Ramond-Ramond field transformation rule obtained
by Bergshoeff et al. is shown as a specific simple example.
We also give some explicit examples of the spinor
representation.

\vskip 6cm

\pagebreak

\parindent=4ex 

\section{Introduction}

The transformation of the fields from the Neveu-Schwarz (NS) sector
under T-duality is well established. The Ramond-Ramond (RR) 
field transformation was first given in \cite{bergshoeff95}.
The authors in \cite{bergshoeff95} got the RR field transformation
by identifying the same RR fields and RR moduli 
in $d=9$ supergravity coming from both ten dimensional
type IIA and type IIB
supergravity theories compactified down to nine dimensions.
Unfortunately this method gives only a 
specific T-duality transformation,
namely Buscher's T-duality transformation
\cite{buscher87}. It is hard to
get the generalized T-duality group $O(d,d)$ transformations
by this method.  Recently, Hassan derived
the RR field transformation under $SO(d,d)$ group
by working on the worldsheet
theory \cite{hassan99}. The RR field transformation
under Buscher's T-duality was also discussed 
by Cveti\v c, L\"{u}, Pope and Stelle 
using the Green-Schwarz formalism \cite{cvetic99}. 
If we compactify the $d=9$ supergravity 
further down to lower dimensions, we know that the 
lower dimensional solution
has $O(d,d,R)$ transformation and how the NS-NS fields
which are assumed to be independent of $d$ coordinates
transform
under this group \cite{sen91}. Therefore, we need to 
find the RR
field transformation under the general $O(d,d,R)$ group.

The RR fields transform as 
the Majorana-Weyl spinors of $SO(d,d)$ \cite{brace98}.
RR fields transforming as the spinors of $O(d,d)$
group is discussed in more detail from the algebraic
decomposition of U duality group in \cite{andrianopoli96}
\footnote{The author thanks S. Ferrara for pointing out
this reference.}.
The spinor representation idea was further developed by
Fukuma, Oota and Tamaka \cite{fukuma99}. It is not
the RR potentials that transform as the 
Majorana-Weyl spinor of $SO(d,d)$; it is the mixed
fields of RR potentials and NS-NS two form that
transform as the Majorana-Weyl spinor of $SO(d,d)$.
However, since the full T-duality group is $O(d,d)$,
we expect to use the Majorana spinor representation
of $O(d,d)$. Note also that $SO(d,d)$ transformations
cannot interchange type IIA and type IIB theories.
In this paper, we use RR fields to construct spinors
of $O(d,d)$ explicitly.
As a simple application
we use the Majorana spinor representation
to show the RR field transformations between
type IIA and type IIB under T-duality. By
using the Majorana spinor and the tensor
representation of $O(d,d,R)$ group, we can get
more general solution generating rules.

We define the RR potentials $C_{p+1}=(1/(p+1)!)\,
C_{\mu_1...\mu_{p+1}}\,dx^{\mu_1}\wedge
\cdots\wedge dx^{\mu_{p+1}}$. Following the definition
given by \cite{fukuma99}, we define the new mixed
fields as
\begin{equation}
\label{newd}
\begin{array}{lcl}
D_0 \equiv C_0, & \quad & D_1 \equiv C_1, \\ 
D_2 \equiv C_2+B_2\wedge C_0, & &
                 D_3 \equiv C_3+ B_2\wedge C_1,\\
D_4 \equiv C_4+{1\over 2} B_2 \wedge C_2
             +{1\over 2} B_2 \wedge B_2 \wedge C_0.
  & & 
\end{array} 
\end{equation}
The RR field strengths are $F=e^{-B}\wedge dD$\cite{fukuma99}
\cite{green96}, here
\begin{equation}
D \equiv \sum_{p=0}^4 D_p,\quad
F \equiv \sum_{p=1}^5 F_p.
\end{equation}
More explicitly, we have
\begin{equation}
\begin{array}{lcl}
F_1\,=\,dD_0, & \quad &  F_2\,=\,dD_1, \\
F_3\,=\,dD_2- B_2\wedge dD_0, & &
                  F_4\,=\,dD_3- B_2\wedge dD_1,\\
F_5\,=\,dD_4- B_2\wedge dD_2
      +{1\over 2} B_2\wedge B_2\wedge dD_0. & & ~
\end{array} 
\end{equation}
We also use the convention
\begin{equation}
\int d^d x \sqrt{-g}|F_p|^2=
\int d^d x {\sqrt{-g}\over p!}
g^{\mu_1\nu_1}\cdots g^{\mu_p\nu_p}F_{\mu_1\mu_p}
F_{\nu_1\nu_p}.
\end{equation}

\section{$d=10$ Type IIA and Type IIB Reduction to $d=9$}

The action of ten dimensional type IIA supergravity can be
written as
\begin{equation}
\label{iia}
\begin{split}
S^{IIA}_{10}=&{1\over 2\kappa^2_{10}}\int d^{10}x \sqrt{-G}
e^{-2\Phi}\left[R(G)+4 G^{MN}\partial_M \Phi
\partial_N \Phi -{1\over 2}
|H_3|^2\right]\\
&-{1\over 4\kappa^2_{10}}\int d^{10}x \sqrt{-G}
\Bigl(|F_2|^2+|F_4|^2\,\Bigr)+{1\over 4\kappa^2_{10}}\int d^{10}x
B_2\wedge dC_3\wedge dC_3,
\end{split}
\end{equation}
where $H_3=dB_2$, $F_2=dC_1=dD_1$,
$F_4=dC_3+H_3\wedge C_1=dD_3-B_2\wedge d D_1$
and the subscript number of a form
denotes the degree of the form.
Now we dimensionally reduce the action (\ref{iia})
to nine dimensions by the vielbein,
\begin{equation}
E^A_M=\begin{pmatrix}
e^a_\mu & e A^{(1)}_\mu\\
0 & e
\end{pmatrix},
\quad
E^M_A=\begin{pmatrix}
e_a^\mu & -e^\nu_aA^{(1)}_\nu\\
0 & e^{-1}
\end{pmatrix}.
\end{equation}
The dimensionally reduced nine dimensional 
action for the NS and R sector is
\begin{equation}
\begin{split}
\label{9sugr}
S_9=&{1\over 2\kappa^2_9}\int d^9x \sqrt{-g}
e^{-2\phi}\biggl[R(g)+4g^{\mu\nu}\partial_\mu \phi\partial_\nu \phi
-e^{-2}g^{\mu\nu}\partial_\mu e\partial_\nu e\\
&-{1\over 2}e^2|F^{(1)}_2|^2
-{1\over 2}e^{-2}|F^{(2)}_2|^2
-{1\over 2}|H_3^{(1)}|^2\biggr]\\
&-{1\over 4\kappa^2_9}\int d^9 x \sqrt{-g}
\Bigl(e|F_2|^2+e^{-1}g^{\mu\nu}\partial_\mu D_x\partial_\nu D_x
+e^{-1}|H^{(2)}_3|^2+e|F_4|^2\Bigr),
\end{split}
\end{equation}
where
\begin{subequations}
\begin{gather}
\label{iia29a}
e^2=G_{xx},\qquad g_{\mu\nu}=G_{\mu\nu}-
G_{xx}A^{(1)}_\mu A^{(1)}_\nu,\\
\label{iia29b}
A^{(1)}_\mu={G_{\mu x}\over G_{xx}},
\qquad A^{(2)}_\mu=B_{\mu x}\\
\label{iia29c}
A_\mu=D_\mu-A^{(1)}_\mu D_x,\quad
F^{i}_{\mu\nu}=\partial_\mu A^{(i)}_\nu
-\partial_\nu A^{(i)}_\mu,\\
\label{iia29d}
B^{(1)}_{\mu\nu}=B_{\mu\nu}+
{1\over 2}A^{(1)}_\mu A^{(2)}_\nu
-{1\over 2}A^{(1)}_\nu A^{(2)}_\mu,\quad 
B^{(2)}_{\mu\nu}=D_{\mu\nu x},\\
\label{iia29e}
\phi=\Phi-\ln\,G_{xx}/4, \qquad 
{\cal D}_{\mu\nu\rho}=D_{\mu\nu\rho},\\
\label{iia29f}
H_3^{(1)}=d B_2^{(1)}-{1\over 2}(A_1^{(1)}\wedge
F_2^{(2)}+A_1^{(2)}\wedge F_2^{(1)}),\\
\label{iia29g}
H_3^{(2)}=d B_2^{(2)}-B_2^{(1)}\wedge d D_x
+{1\over 2}A_1^{(2)} \wedge A_1^{(1)}\wedge d D_x
-A_1^{(2)} \wedge (F_2+ F_2^{(1)} D_x),\\
\label{iia29h}
F_4=d {\cal D}_3-B_2^{(1)}\wedge d D_1
+{1\over 2}A_1^{(1)}\wedge A_1^{(2)} \wedge d D_1
+H_3^{(2)} \wedge A_1^{(1)},
\end{gather}
\end{subequations}
and $x$ is the compactified coordinate. Here we 
follow the general
prescription of dimensional reduction given in
\cite{maharana92}. For example, the lower dimensional
field strength comes from the higher dimensional
field strength as $H^{(1)}_{\mu\nu\rho}=e^a_\mu e^b_\nu
e^c_\rho E^M_a E^N_b E^P_c H_{MNP}$.
The action (\ref{9sugr}) can be obtained from the type
IIB supergravity in ten dimensions also if we use the following
vielbein for the IIB theory \cite{bergshoeff95}
\begin{equation}
{\mathst E}^A_M=\begin{pmatrix}
e^a_\mu & e^{-1} A^{(2)}_\mu \\
0 & e^{-1}
\end{pmatrix},
\quad
{\mathst E}^M_A=\begin{pmatrix}
e_a^\mu & -e^\nu_a A^{(2)}_\nu\\
0 & e
\end{pmatrix},
\end{equation}
together with the following definitions,
\begin{subequations}
\begin{gather}
\label{iib29a}
e^{-2}={\mathst G}_{xx},\qquad g_{\mu\nu}
={\mathst G}_{\mu\nu}-
{\mathst G}_{xx}A^{(2)}_\mu A^{(2)}_\nu,\\
\label{iib29b}
A^{(1)}_\mu={\mathst B}_{\mu x}, \qquad
A^{(2)}_\mu={{\mathst G}_{\mu x}\over {\mathst G}_{xx}},\qquad
D=D_x,\\
\label{iib29c}
A_\mu=D_{\mu x}-{\mathst B}_{\mu x} D
=D_{\mu x}-A^{(1)}_\mu D,\\
\label{iib29d}
B^{(1)}_{\mu\nu}={\mathst B}_{\mu\nu}-
{1\over 2}A^{(1)}_\mu A^{(2)}_\nu
+{1\over 2}A^{(1)}_\nu A^{(2)}_\mu,\qquad 
B^{(2)}_{\mu\nu}=D_{\mu\nu},\\
\label{iib29e}
\phi={\hat \Phi}-\ln\,{\mathst G}_{xx}/4, \qquad 
{\cal D}_{\mu\nu\rho}=D_{\mu\nu\rho x}.
\end{gather}
\end{subequations}
The type IIB ten dimensional supergravity action we
use is
\begin{equation}
\label{iib}
\begin{split}
S^{IIB}_{10}=&{1\over 2\kappa^2_{10}}\int d^{10}x 
\sqrt{-{\mathst G}}
e^{-2\hat \Phi}\left[R({\mathst G})+
4 {\mathst G}^{MN}\partial_M \hat \Phi
\partial_N \hat \Phi -{1\over 2}
|{\mathst H}_3|^2\right]\\
&-{1\over 4\kappa^2_{10}}\int d^{10}x \sqrt{-{\mathst G}}
\left(|F_1|^2+|F_3|^2+{1\over 2}|F_5|^2\,\right)
+{1\over 4\kappa^2_{10}}\int d^{10}x
{\mathst B}_2\wedge dC_4\wedge dC_2,
\end{split}
\end{equation}
together with the self dual constraint on $F_5$.
Now we can get Buscher's T-duality transformations
\cite{buscher87} from Eqs. (\ref{iia29a})-(\ref{iia29e})
and Eqs. (\ref{iib29a})-(\ref{iib29e}) as follows
\begin{subequations}
\begin{gather}
\label{factdula}
{\tilde g}_{xx}={1\over g_{xx}},\quad
{\tilde g}_{\mu x}={B_{\mu x}\over g_{xx}},\quad
{\tilde g}_{\mu\nu}=g_{\mu\nu}-{g_{\mu x}g_{\nu x}-B_{\mu x}B_{\nu x}
\over g_{xx}},\\
\label{factduld}
{\tilde B}_{\mu x}={g_{\mu x}\over g_{xx}},\quad
{\tilde B}_{\mu\nu}=B_{\mu\nu}-{B_{\mu x}g_{\nu x}
-B_{\nu x}g_{\mu x} \over g_{xx}},\\
\label{factdulf}
{\tilde \phi}=\phi-{1\over 2}\ln g_{xx},\\
\label{factdulg}
{\tilde D}_x=D,\quad {\tilde D}_\mu=D_{\mu x},\quad
{\tilde D}_{\mu\nu x}=D_{\mu\nu},\quad
{\tilde D}_{\mu\nu\rho}=D_{\mu\nu\rho x}.
\end{gather}
\end{subequations}
From the above transformation rules (\ref{factdula})-
(\ref{factdulg}), we have the following transformations 
in terms of the original RR potentials,
\begin{subequations}
\begin{gather}
\label{rra}
{\tilde C}_x=C,\quad 
{\tilde C}_\mu=C_{\mu x}+B_{\mu x}C,\quad
{\tilde C}_{\mu\nu x}=C_{\mu\nu}+{g_{\mu x}C_{\nu x}
-g_{\nu x}C_{\mu x}\over g_{xx}},\\
\label{rrb}
{\tilde C}_{\mu\nu\rho}=C_{\mu\nu\rho x}-{3\over 2}
B_{[\mu\nu}C_{\rho] x}-{3\over 2}B_{x[\mu}C_{\nu\rho]}
-{6g_{x[\mu}B_{\nu|x|}C_{\rho]x}\over g_{xx}}.
\end{gather}
\end{subequations}
In general we should consider the 
$O(d,d,R)$ transformations. The group element
$\Omega$ of $O(d,d,R)$ satisfies 
\begin{equation}
\label{4dL}
\Omega^T J \Omega = J ,\qquad 
J =\begin{pmatrix}
0 & 1\!\!1_d\\
1\!\!1_d & 0
\end{pmatrix}.
\end{equation}
If we put the NS sector fields in a $2d$ by $2d$ matrix
\begin{equation}
\label{modulthree}
M=\begin{pmatrix}
G^{-1} & -G^{-1}B \\
B G^{-1} & G -B G^{-1} B
\end{pmatrix}=\begin{pmatrix}
1\!\!1 &0\\
B&1\!\!1
\end{pmatrix}
\begin{pmatrix}
G^{-1}&0\\
0&G
\end{pmatrix}
\begin{pmatrix}
1\!\!1 &-B\\
0&1\!\!1
\end{pmatrix},
\end{equation}
where $G=[G_{ij}]$ and $B=[B_{ij}]$ are $d\times d$ matrices,
$i$ and $j$ run over the compactified or independent
$d$ coordinates. Let
\begin{gather}
\label{defa}
A_{\mu m}^{(1)}=G_{\mu m},\quad A^{(1)m}_\mu=G^{mn} A_{\mu n}^{(1)},\\
\label{defb}
A^{(2)}_{\mu m} = B_{\mu m} + B_{m n}
A^{(1) n}_{\mu},\quad
{\cal A}^i_{\mu}
=\begin{pmatrix} 
A^{(1)m}_{\mu}\\
A^{(2)}_{\mu\,m}
\end{pmatrix},\\
\label{defc}
g_{\mu\nu}=G_{\mu\nu}-G_{mn} A^{(1)m}_\mu A^{(1)n}_{\nu},\\
\label{defd}
\phi=\Phi-{1\over 4}\ln\, {\rm det}(G_{mn}),\\
\label{defe}
B_{\mu \nu} = \hat B_{\mu \nu} + {1 \over 2} A^{(1) m}_{\mu}
A^{(2)}_{\nu m} - {1 \over 2} A^{(1) m}_{\nu} A^{(2)}_{\mu
m} - A^{(1) m}_{\mu} B_{m n} A^{(1) n}_\nu,
\end{gather}
where $\Phi$, $G_{\mu m}$, $G_{\mu\nu}$, 
$G_{mn}$, $\hat B_{\mu\nu}$, $B_{\mu m}$ and $B_{mn}$
are the original NS fields.
The $O(d,d)$ transformations for the NS fields are
\begin{equation}
M \to \Omega M \Omega^T ,\ \ \ {\cal A}^i_{\mu} \to \Omega_{ij}
{\cal A}^j_{\mu}, \ \ \ g_{\mu\nu} \to g_{\mu\nu}, \ \ \
\phi \to \phi, \ \ \ B_{\mu\nu} \to B_{\mu\nu}.
\label{tdual}
\end{equation}

\section{Spinor Representation}

In this section, we will show that the RR fields transform
as the Majorana spinors.
We can write the general group element $\Omega$ of $O(d,d,R)$
as
\begin{equation}
\Omega=\begin{pmatrix}
{\mathst A} & {\mathst B}\\
{\mathst C} & {\mathst D}
\end{pmatrix},
\end{equation}
with ${\mathst A}{\mathst B}^T+{\mathst B}{\mathst A}^T=
{\mathst C}{\mathst D}^T+{\mathst D}{\mathst C}^T=0$,
${\mathst A}{\mathst D}^T+{\mathst B}{\mathst C}^T=
{\mathst C}{\mathst B}^T+{\mathst D}{\mathst A}^T=1$,
${\mathst A}$, ${\mathst B}$, ${\mathst C}$ and
${\mathst D}$ are $d\times d$ matrices. We can also
show that ${\mathst D}={\mathst C}{\mathst A}^{-1}
{\mathst B}+({\mathst A}^{-1})^T$. The $O(d,d,R)$
group can be generated by the following three
matrices \cite{giveon94}
\begin{equation}
\Lambda_C=\begin{pmatrix}
1\!\!1& 0\\
C & 1\!\!1
\end{pmatrix},\quad
\Lambda_R=\begin{pmatrix}
(R^T)^{-1} & 0\\
0 & R
\end{pmatrix},\quad
\Lambda_i=\begin{pmatrix}
-1\!\!1 + e_i & e_i\\
e_i & -1\!\!1 +e_i
\end{pmatrix},\quad
(e_i)_{jk}=\delta_{ij}\delta_{jk},
\end{equation}
where $C^T=-C$, $R\in GL(d,R)$ and $i$, $j$, and $k=1$ ,
$\dots$, $d$. The action of $\Lambda_C$ shifts
the NS two-form by the matrix $C$. Under the action
of $\Lambda_R$, $G\rightarrow RGR^T$, $B\rightarrow
RBR^T$.
For the group $O(d,d,Z)$, we need to restrict
the matrix elements to be integers.

The Dirac matrices satisfy $\{\Gamma_r,\ \Gamma_s\}=2J_{rs}$
with $r$ and $s=1$, $\dots$, $2d$. Let
\begin{equation}
\label{crean}
a_i={\Gamma_{d+i}\over \sqrt{2}},\quad
a_i^\dag={\Gamma_i\over \sqrt{2}},\quad
i=1,\ \dots,\ d.
\end{equation}
Then we have $\{a_i$, $a_j^\dag\}=\delta_{ij}
1\!\!1$, $\{a_i$, $a_j\}=\{a_i^\dag$, $a_j^\dag\}=0$.
Define the vacuum to be $a_i|0\rangle=0$, we
can get the representation (Fock) space as
\begin{equation}
\label{state}
|\alpha\rangle=(a_1^\dag)^{i_1}\cdots
(a_d^\dag)^{i_d} |0\rangle, \quad 
i_1,\ \dots,\ i_d=0\ {\rm or}\ 1.
\end{equation}
The spinor representation of the $O(d,d)$ group is given by
\begin{equation}
\label{spinrep}
S(\Omega)\Gamma_s S(\Omega)^{-1}=\sum_r \Gamma_r\Omega^r\,_s.
\end{equation}
For convenience we can define the operator corresponding to
a matrix $\Omega$ as
\begin{equation}
\label{operator}
{\mathbold \Omega}\Gamma_s = \sum_r\Gamma_r\Omega^r\,_s
{\mathbold \Omega},\quad
{\mathbold \Omega}|\beta\rangle=\sum_\alpha 
|\alpha\rangle S_{\alpha\beta}(\Omega).
\end{equation}
The operators for the three generating 
matrices are \cite{brace98}\cite{fukuma99}
\begin{gather}
{\mathbold \Lambda}_C=\exp\left({1\over 2}
C_{ij}a_i a_j\right),\quad
{\mathbold \Lambda}_i=\pm (a_i+a_i^\dag),\\
{\mathbold \Lambda}_R=({\rm det}R)^{-1/2}\,\exp
\left(a_i A_i\,^j a_j^\dag\right),\quad
R=R_i\,^j=\exp(A_i\,^j),
\end{gather}
where the repeated indices are summed.
We choose $+$ sign for the ${\mathbold \Lambda}_i$ operator.
The new mixed $D$ fields form a spinor as follows: for $d=1$,
\begin{gather*}
\chi_\alpha=(D,\ D_x),\quad \chi_{\mu\alpha}=(D_\mu,\ D_{\mu x}),\\
\chi_{\mu\nu\alpha}=(D_{\mu\nu},\ D_{\mu\nu x}),\quad
\chi_{\mu\nu\rho\alpha}=(D_{\mu\nu\rho},\ D_{\mu\nu\rho x}),\\
\cdots,
\end{gather*}
with $|\alpha\rangle=(|0\rangle,\ a^\dag|0\rangle)$; for $d=2$,
\begin{gather*}
\chi_\alpha=(D,\ D_x,\ D_y,\ D_{yx}),\\
\chi_{\mu\alpha}=(D_\mu,\ D_{\mu x},\ D_{\mu y},
\ D_{\mu yx}),\\
\chi_{\mu\nu\alpha}=(D_{\mu\nu},\ D_{\mu\nu x},
\ D_{\mu\nu y},\ D_{\mu\nu yx}),\\
\cdots,
\end{gather*}
with $|\alpha\rangle=(|0\rangle,\ a_x^\dag|0\rangle,\ 
a_y^\dag|0\rangle,\ a_x^\dag a_y^\dag|0\rangle)$ and so on.
The fields $\chi$ transform as 
\begin{equation}
\label{spintrans}
|{\tilde \chi}_{\mu_1\dots\mu_p\alpha}\rangle
=\sum_\beta S^{-1}(\Omega^T)_{\alpha\beta}
|{\tilde \chi}_{\mu_1\dots\mu_p\beta}\rangle.
\end{equation}
For instance, the spinor representation matrix
of $O(1,1)$  for $\Lambda_i$ is
\begin{equation}
\label{rdualex}
S\left((\Lambda^T)^{-1}\right)=S(\Lambda)
=\Lambda=\begin{pmatrix}
0 & 1\\
1 & 0
\end{pmatrix}.
\end{equation}
From the spinor matrix (\ref{rdualex}), it is easy to
get Buscher's T-duality transformations
(\ref{factdula})-(\ref{factdulg}), (\ref{rra})
and (\ref{rrb}) by combining Eqs. (\ref{tdual}) and
(\ref{spintrans}). The spinor representation of $SO(1,1)$
for $\Lambda_i\Lambda_j$ is
\begin{equation}
\label{rdualex1}
S(\Lambda^2)=\Lambda^2=\begin{pmatrix}
1 & 0\\
0 & 1
\end{pmatrix}.
\end{equation}
This is a trivial identity transformation. Furthermore
it gives a Majorana-Weyl spinor representation.

\section{More Examples}

In order to discuss the solution generating transformations,
we focus on $O(d)\otimes O(d)$ group in this section.
We embed the $O(d)$ matrices $R$ and $S$
into $O(d,d)$ matrix $\Omega$.
Because the metric $J$ of $O(d,d)$ 
is rotated from the diagonal metric $\eta$ by
$$J={\mathst R}\eta {\mathst R},\quad
\eta=\begin{pmatrix}
-1\!\!1 & 0\\
0 & 1\!\!1
\end{pmatrix},\quad
{\mathst R}={\sqrt{2}\over 2}
\begin{pmatrix}
-1\!\!1 & 1\!\!1\\
1\!\!1  & 1\!\!1
\end{pmatrix},$$
so
$$\Omega={\mathst R}^{-1}\begin{pmatrix}
S & 0\\
0 & R
\end{pmatrix}{\mathst R}={1\over 2}\begin{pmatrix}
R+S & R-S\\
R-S & R+S
\end{pmatrix}.$$
Note that $\Omega$ is also an element of $O(2d)$, so
$(\Omega^T)^{-1}=\Omega$.

For example, if we take $R=-1\!\!1 +2 e_i$, $S=-1\!\!1$,
then we recover the T-duality $\Lambda_i$ discussed before.
If we choose \cite{hassan99}
$$S=1\!\!1,\qquad R=\begin{pmatrix}
\cos\theta & \sin\theta\\
-\sin\theta & \cos\theta
\end{pmatrix},$$
then we have the spinor representation
\begin{equation}
\label{soluex1b}
S(\Omega)=\begin{pmatrix}
\cos{\theta\over 2} & 0 & 0 &-\sin{\theta\over 2}\\
0 & \cos{\theta\over 2} & \sin{\theta\over 2}& 0 \\
0 & -\sin{\theta\over 2} & \cos{\theta\over 2} & 0 \\
\sin{\theta\over 2} & 0 & 0 & \cos{\theta\over 2}
\end{pmatrix}.
\end{equation}
For flat background with zero $B$ field, RR
fields transform the same way as the $D$ fields.
This result is consistent with that obtained
in \cite{hassan99}.

If one of the coordinate is timelike, we have
\begin{equation}
\label{gentec2} 
\Omega={1\over 2}\begin{pmatrix}
\eta(S+R)\eta & \eta(R-S)\\
(R-S)\eta & S+R
\end{pmatrix},\quad
{\mathst R}={1\over \sqrt{2}}\begin{pmatrix}
-\eta& 1\!\!1\\
\eta & 1\!\!1
\end{pmatrix},
\end{equation}
here $\eta$ is the Minkowski metric, 
$S$ and $R$ are $O(d-1,1)$ matrices satisfying
$S\eta S^T=\eta$ and
$R\eta R^T=\eta$.
For example, 
$$ R=S=\begin{pmatrix}
\cosh\alpha & \sinh \alpha\\
\sinh\alpha & \cosh\alpha
\end{pmatrix}$$
generate the boost transformation along $t$-$x$ 
coordinates,
\begin{equation}
\label{soluexbb}
S^{-1}(\Omega^T_b)=\begin{pmatrix}
1 & 0 & 0 & 0\\
0 & \cosh\alpha & \sinh\alpha & 0\\
0 & \sinh\alpha &\cosh\alpha & 0 \\
0 & 0 & 0 & 1
\end{pmatrix}.
\end{equation}
More explicitly, the boost transformation for the RR
field and $B$ field is
\begin{subequations}
\begin{gather}
\label{soluexb5}
{\tilde B}_{\mu t}=B_{\mu t}\cosh\alpha
+B_{\mu x}\sinh\alpha,\quad 
{\tilde B}_{\mu x}=B_{\mu t}\sinh\alpha
+B_{\mu x}\cosh\alpha,\\
\label{soluexb6}
{\tilde C}_{\mu\dots\nu t}=C_{\mu\dots\nu t}\cosh\alpha
+C_{\mu\dots\nu x}\sinh\alpha,\quad 
{\tilde C}_{\mu\dots\nu x}=C_{\mu\dots\nu t}\sinh\alpha
+C_{\mu\dots\nu x}\cosh\alpha,\\
\label{soluexb7}
{\tilde B}_{tx}=B_{tx},\quad
{\tilde C}_{\mu\dots\nu tx}=C_{\mu\dots\nu tx},\quad
{\tilde B}_{\mu\nu}=B_{\mu\nu},\quad
{\tilde C}_{\mu\dots\nu}=C_{\mu\dots\nu}.
\end{gather}
\end{subequations}
Finally let us choose\cite{sen91}
\begin{equation}
\label{soluex4}
S=\begin{pmatrix}
\cosh \alpha & -\sinh\alpha\\
-\sinh\alpha & \cosh\alpha
\end{pmatrix},
\quad R=\begin{pmatrix}
\cosh \alpha & \sinh\alpha\\
\sinh\alpha & \cosh\alpha
\end{pmatrix}.
\end{equation}
In this case, we have 
\begin{equation}
\label{soluex4b}
S^{-1}(\Omega^T_s)=\begin{pmatrix}
\cosh\alpha & 0 & 0 &\sinh\alpha\\
0 & 1 & 0 & 0\\
0 & 0 & 1 & 0\\
\sinh\alpha & 0 & 0 &\cosh\alpha
\end{pmatrix}.
\end{equation}

For the background $B_{\mu\nu}=0$, $g_{11}=1$ and
$g_{01}=0$, the transformations of NS-NS
and RR fields are
\begin{subequations}
\begin{gather}
\label{gtransfa}
{\tilde g}_{00}={g_{00}\over 1+(1+g_{00})\sinh^2\alpha},\\
\label{gtransfb}
{\tilde g}_{11}={1\over 1+(1+g_{00})\sinh^2\alpha},\\
\label{atransfa}
{\tilde B}_{01}={(1+g_{00})\sinh 2\alpha\over 
2[1+(1+g_{00})\sinh^2\alpha]},
\end{gather}
\begin{gather}
\label{gtransfc}
{\tilde g}_{\mu 0}={g_{\mu 0} \cosh\alpha \over
1+(1+g_{00})\sinh^2\alpha},\\
\label{gtransfd}
{\tilde g}_{\mu 1}={g_{\mu 1} \cosh\alpha \over
1+(1+g_{00})\sinh^2\alpha},\\
\label{atransfb}
{\tilde B}_{\mu 0}={-g_{00}g_{\mu 1}\sinh\alpha
\over 1+(1+g_{00})\sinh^2\alpha},\\
\label{atransff}
{\tilde B}_{\mu 1}={g_{\mu 0}\sinh\alpha\over 
1+(1+g_{00})\sinh^2\alpha},
\end{gather}
\begin{gather}
\label{gtransfe}
{\tilde g}_{\mu\nu}=g_{\mu\nu}-{(g_{\mu 0}
g_{\nu 0}+g_{00}g_{\mu 1}g_{\nu 1})\sinh^2\alpha
\over
1+(1+g_{00})\sinh^2\alpha},\\
\label{atransfc}
{\tilde B}_{\mu\nu}={(g_{\mu 0}g_{\nu 1}
-g_{\mu 1}g_{\nu 0})\sinh\alpha\,\cosh\alpha
\over
1+(1+g_{00})\sinh^2\alpha},
\end{gather}
\begin{gather}
\label{ctransfa}
{\tilde C}=C\cosh\alpha-C_{01}\sinh\alpha,\\
\label{ctransfb}
{\tilde C}_0=C_0,\quad
{\tilde C}_1=C_1,\quad
{\tilde C}_\mu=C_\mu\cosh\alpha-C_{\mu 01}\sinh\alpha,\\
\label{ctransfc}
\begin{split}
{\tilde C}_{01}=& {C_{01}[1+2(1+g_{00})\sinh^2\alpha]\cosh\alpha
\over 1+(1+g_{00})\sinh^2\alpha}\\
&-{C[1+(1+g_{00})(\sinh^2\alpha+\cosh^2\alpha)]\sinh\alpha
\over 1+(1+g_{00})\sinh^2\alpha},
\end{split}
\end{gather}
\begin{gather}
\label{ctransfd}
{\tilde C}_{\mu 0}=C_{\mu 0}+{g_{00}g_{\mu 1}\sinh\alpha
(C \cosh\alpha -C_{01}\sinh\alpha)\over
1+(1+g_{00})\sinh^2\alpha},\\
\label{ctransfe}
{\tilde C}_{\mu 1}=C_{\mu 1}-{C g_{\mu 0}
\sinh\alpha\cosh\alpha\over 
1+(1+g_{00})\sinh^2\alpha}
+{C_{01}g_{\mu 0}\sinh^2\alpha\over 
1+(1+g_{00})\sinh^2\alpha},\\
\label{ctransff}
\begin{split}
{\tilde C}_{\mu\nu}=& C_{\mu\nu}\cosh\alpha
-C_{\mu\nu 01}\sinh\alpha\\
&+{(C_{01}\sinh\alpha-C\cosh\alpha)(g_{\mu 0}g_{\nu 1}
-g_{\mu 1}g_{\nu 0})\sinh 2\alpha\over
2[1+(1+g_{00})\sinh^2\alpha]},
\end{split}\\
\label{phitransfa}
e^{-2{\tilde \phi}}=e^{-2\phi}[1+(1+g_{00})\sinh^2\alpha].
\end{gather}
\end{subequations}

\section{Discussion}

The RR field transformations are very simple in terms
of the new mixed fields $D$. It is very easy to
see the RR field transformations from the spinor
representations. 
For any group element $\Omega\in O(d,d)$, we can
get the spinor representation $S(\Omega)$ from
Eq (\ref{spinrep}) or Eq. (\ref{operator}).
We can introduce higher degree
potentials and field strengths with some constraints
as shown in \cite{fukuma99}. With the extra potentials,
the action for the RR and Chern-Simons terms
can be written in a simple way. This may suggest
that the $D$ fields are the natural RR potentials.
We can apply the transformation Eqs. (\ref{tdual}) for the
NS-NS fields and the transformation Eqs. (\ref{spintrans})
for the RR fields to get more general solution generating
rules.  

\bigskip

{\noindent Note Added:} In the second version of paper
\cite{hassan99}, Hassan gave a general transformation
of $D$ field by spinor representation and discussed
the equivalence of the RR field transformations between
his supersymmetry method and the spinor representation.

\end{document}